\documentclass[sigconf]{acmart}
\usepackage{subfigure}
\usepackage{bm}
\usepackage{xspace}
\usepackage{multirow}
\usepackage{algorithm}
\usepackage{algorithmic}
\usepackage{array}
\usepackage[normalem]{ulem}
\usepackage{epstopdf}
\usepackage{setspace}
\usepackage{pifont}
\usepackage[skip=0pt]{caption}
\usepackage{subfigure}
\usepackage{xcolor}
\usepackage{mathrsfs}

\newcommand{\paratitle}[1]{\vspace{1.5ex}\noindent\textbf{#1}}
\newcommand{\ie}{\emph{i.e.,}\xspace}

\newcommand{\eg}{\emph{e.g.,}\xspace}

\newcommand{\ignore}[1]{}

\newcommand{\changed}[1]{\textcolor{blue}{#1}}

\setcopyright{rightsretained}

\acmYear{2018}
\copyrightyear{2018}

\title{KB4Rec: A Dataset for Linking Knowledge Bases with Recommender Systems}
\author{Wayne Xin Zhao, Gaole He, Hongjian Dou, Jin Huang, Siqi Ouyang and Ji-Rong Wen}
\affiliation{%
  \institution{School of Information, Renmin University of China}
}
\email{{batmanfly,ouyangsiqi0726}@gmail.com, {hegaole, hongjiandou, jin.huang, jrwen}@ruc.edu.cn}

\begin{document}

\begin{abstract}
To develop a knowledge-aware recommender system, a key data problem is how we can obtain rich and structured knowledge information for recommender system (RS) items. Existing datasets or methods either use side information from original recommender systems (containing very few kinds of useful information) or utilize private knowledge base (KB).  In this paper, we present  the first public linked KB dataset for recommender systems, named \emph{KB4Rec v1.0}, which has linked three widely used RS datasets with the popular KB Freebase.
Based on our linked dataset, we first preform some interesting qualitative analysis experiments, in which we discuss the effect of two important factors (\ie popularity and recency) on whether a RS item can be linked
to a KB entity. Finally, we present the comparison of several knowledge-aware recommendation algorithms on our linked dataset.
\end{abstract}

%
%
\ignore{
}



\maketitle

\section{Introduction}
With the rapid development of Web techniques, various kinds of side information has become available in recommender systems (RS). In an early stage, such context information is usually unstructured, and its availability is limited to specific data domains or platforms~\cite{Harper2016The, Schedl2016The, He2016Ups}.
Recently, more and more efforts have been made by both research and industry communities for structuring world knowledge or domain facts  in a variety of data domains. One of the most typical organization forms is \emph{knowledge base (KB)}~\cite{KBE-survey}.
KBs provide a general and unified way to organize and relate information entities, which have been shown to be useful in many applications. Specially, KBs have  been used in RSs~\cite{KDD-2016-Zhang,HuangZDWC18}, usually called \emph{knowledge-aware recommendation}.

To develop a knowledge-aware recommender system, a key data problem is how we can obtain rich and structured knowledge information for RS items.
Overall, there are two main solutions from existing studies.
First, side information is collected from the RS platform~\cite{Harper2016The, Schedl2016The, He2016Ups}, and several studies further construct tiny and simple KB-like knowledge structure~\cite{Yu-2014-WSDM}. The number of attributes or relations is usually limited, and much useful knowledge information has not been considered. Second, several works propose to link RS with private KBs~\cite{KDD-2016-Zhang}.
The linkage results are not publicly available.

To address the need for the linked dataset of RS and KBs, we present a public linked KB dataset for recommender systems, named \emph{KB4Rec} v1.0, freely available at \changed{\url{https://github.com/RUCDM/KB4Rec}}.
Our basic idea is to heuristically link items from RSs with entities from a public large-scale  KB\footnote{
We use the terms of ``items" and ``entities" respectively for RSs and KBs. }. On the RS side, we select three widely used datasets (\ie MovieLens~\cite{Harper2016The}, LFM-1b~\cite{Schedl2016The} and Amazon book~\cite{He2016Ups}) covering three different data domains, namely movie, music and book; on the KB side, we select the well-known Freebase~\cite{freebase}.  We try to maximize the applicability of our linked dataset by selecting very popular RS datasets and KBs. 
Specially, we are also aware of some closely related studies~\cite{dbpedia,noia-tist-2016}, which also aim to link RS items with KB entities.
While, our focus is on the Freebase, which is now widely used in many NLP or related domains~\cite{KBE-survey}.

In our KB4Rec v1.0 dataset, we organized the linkage results by linked ID pairs, which consists of a RS item ID and a KB entity ID. We do not share the original datasets, since they are  maintained by original researchers or publishers. All the IDs are inner values from the original datasets. Once such a linkage has been accomplished, it is able to reuse existing large-scale KB data for RSs.
For example, the movie of ``Avatar" from MovieLens dataset~\cite{Harper2016The} has a corresponding entity entry in Freebase, and we are able to obtain its attribute information by reading out all its associated relation triples in Freebase.
Based on the linked dataset, we first preform some interesting qualitative analysis experiments, in which we discuss the effect of two important factors (\ie popularity and recency) on whether a RS item can be linked
to a KB entity. Finally, we present the comparison of several knowledge-aware recommendation algorithms on our linked dataset.



\section{Existing  Datasets and Methods}
In this section, we briefly review the related datasets and methods.

Early knowledge-aware recommendation algorithms are also called context-aware recommendation algorithms, in which  the side information from the original RS platform is considered as context data.
For example, social network information of Epinions dataset is utilized in \cite{Jamali-2010-RecSys, Ma-2009-SIGIR}, POI property information of Yelp dataset is utilized in \cite{Gao-2015-AAAI},  movie attribute information of MovieLens dataset is utilized in \cite{Yu-2014-WSDM} and user profile information of microblogging dataset has been utilized in \cite{Zhao-KDD-2014}.
These datasets usually contain very few kinds of side information, and the relation between different kinds of side information is ignored.

To make such side information more structured, Heterogeneous Information Networks (HIN) have been proposed as a general technique for modeling information networks~\cite{Sun-2012-KDD}. In HINs, we can effectively learn underlying relation patterns (called \emph{meta-path}) and organize side information via meta-path-based representations. For example, HIN-based recommendation have been applied to solve PER~\cite{Yu-2014-WSDM} and MCRec~\cite{Hu-2018-KDD}.
HIN based algorithms usually rely on graph search algorithms, which is difficult to deal with large-scale relation pattern finding.

More recently, KBs have become a popular kind of data resources to store and organize world knowledge or domain facts. Many studies  have been proposed~\cite{KBE-survey} for the construction, inference and applications of KBs.
Specially, several pioneering studies try to leverage existing KB information for improving the recommendation performance~\cite{WWW-2018-Wang,KDD-2016-Zhang,Wang-2018-RippleNetwork}.
They apply a heuristic method for linking RS items with KB entities.
In these studies, they use a private KB for linkage, which cannot be obtained publicly.

Specially, we are also aware of some closely related studies, including~\cite{dbpedia,noia-tist-2016}, which also aim to link RS items with KB entities.
While, our focus is on the Freebase, which is now widely used in many NLP or related domains~\cite{KBE-survey}.

\section{Linked Dataset Construction}
\label{sec:data-collection}

In our work, we need to prepare two kinds of datasets, namely RS and KB data.
 Next, we first give the detailed descriptions of
the original datasets, and then discuss the linkage method.


\paratitle{RS Datasets}. We consider three popular RS datasets for linkage, namely MovieLens, LFM-1b and Amazon book, which covers the three domains of movie, music and book respectively.

 (1) \emph{MovieLens } dataset~\cite{Harper2016The} describes users' preferences on movies.
A preference record takes the form  $\langle$user, item, rating, timestamp$\rangle$, indicating
the rating score of a user for a movie at some time.
There have been four MovieLens datasets released, known as $100K$, $1M$, $10M$, and $20M$,
reflecting the approximate number of ratings in each dataset. We select the largest
MovieLens $20M$ for linkage.

(2) \emph{LFM-1b} dataset~\cite{Schedl2016The} describes users' interaction records on music.
It provides information including artists, albums,
tracks, and users, as well as individual listening events. It records the the listening events of a user on songs, but does not contain rating information.

(3) \emph{Amazon book}  dataset~\cite{He2016Ups} describes users' preferences on book products with the  data form of  $\langle$user, item, rating, timestamp$\rangle$.
The  dataset is very sparse, containing 22 million ratings
from 8 million users across nearly 23 million items.

In the three RS datasets, we several kinds of side information such as item titles (all), IMDB ID (movie), writer (book) and artist (music). We utilize such side information for subsequent KB linkage.  

\paratitle{KB Dataset}. We adopt the large-scale pubic KB \emph{Freebase}. Freebase~\cite{freebase} is a KG announced by Metaweb Technologies, Inc. in 2007 and was acquired by Google Inc. on July 16, 2010. Freebase stores facts by triples of the form $\langle$head, relation, tail$\rangle$. 
Since Freebase shut down its services on August 31, 2016, we use its latest public version. We select Freebase because it has been widely applied in the research communities~\cite{KBE-survey}.

\begin{table}[!htbp]
  \centering
  \caption{Statistics of the linkage results. The three domains 
  correspond to the RS datasets of MovieLens $20M$, LFM-1b and Amazon book, respectively.}
  \label{tab-link-data}%
    \begin{tabular}{c||cccc}
\hline
   Datasets & \#Items & \#Linked-Items & \#Users & \#Interactions \\
   \hline
\hline
  Movie &27,279& 25,982 & 138,493 & 20,000,263 \\
  Music & 6,479,700 & 1,254,923 & 120,317 & 1,021,931,544  \\
  Book & 2,330,066& 109,671 &8,026,324 & 22,507,155 \\
    \hline
    \end{tabular}%
\end{table}%

\paratitle{RS to KB Linkage}. 
With an offline Freebase search API, we retrieve KB entities
 with item titles as queries.
 If no KB entity with the same title was returned, we say the RS item is \emph{rejected} in the linkage process.
 If at least one KB entity with the same title was returned, we further incorporate
 one kind of side information as a refined constraint for accurate linkage: \emph{IMDB ID}, \emph{artist name} and \emph{writer name} are used for the three domains of movie, music and book respectively. We find only a small number (about one thousand for each domain) of RS items can not be accurately linked or rejected via the above procedure, and simply discard them.
During the linkage process, we deal with  several problems that will affect the results of string match algorithms, \eg lowercase,  abbreviation, and the order of family/given names.  Since the LFM-1b dataset is extremely large,
we remove all the musics with fewer than ten listening events. Even after filtering, it still contains about 6.5 million musics.

\paratitle{Basic Statistics}.  We summarize the basic statistics of the three linked datasets in the second column of Table~\ref{tab-link-data}. It can be observed that for the MovieLens $20M$ dataset, we have a very high linkage ratio: about 95.2\% items can be accurately linked to a KB entity. For LFM-1b dataset, the linkage ratio is 19.4\%. But, the linkage ratio for the book domain is very low, about 4.7\%.
A possible explanation is  that MovieLens $20M$ dataset contains fewer items than the other two datasets, which are ready refined by original releasers.
Besides, we speculate that there may exist domain bias in the construction of Freebase.
Although the linkage ratios for the latter two datasets are not high, the absolute numbers of linked items are large.
Such a linked dataset is feasible for research-purpose studies.


\paratitle{Shared Datasets}. We name the above linked linked KB dataset for recommender systems as \emph{KB4Rec} v1.0, freely available at \changed{\url{https://github.com/RUCDM/KB4Rec}}.
In our KB4Rec v1.0 dataset, we organized the linkage results by linked ID pairs, which consists of a RS item ID and a KB entity ID. All the IDs are inner values from the original datasets.
We have 25,982, 1,254,923, and 109,671 linked ID pairs for MovieLens $20M$, LFM-1b and Amazon book respectively.


\section{Linkage Analysis}
\label{sec:data-analysis}

Previously, we have shown the linkage ratios for different datasets. We find that a considerable amount of RS items  can not be linked to KB entities. It is interesting to study what factors will affect the linkage ratio. We consider two kinds of factors for analysis.

\begin{figure*}[htbp]
\centering
\subfigure[Popularity.]{
\begin{minipage}[t]{0.25\linewidth}
\centering
\includegraphics[width=1.60in]{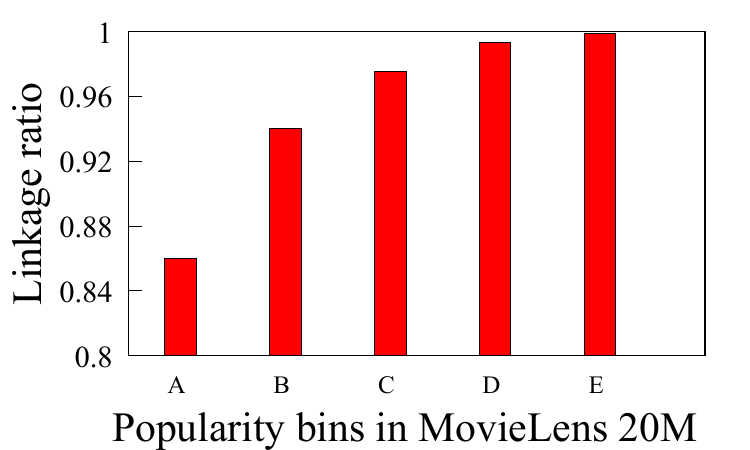}
\end{minipage}%
\begin{minipage}[t]{0.25\linewidth}
\centering
\includegraphics[width=1.60in]{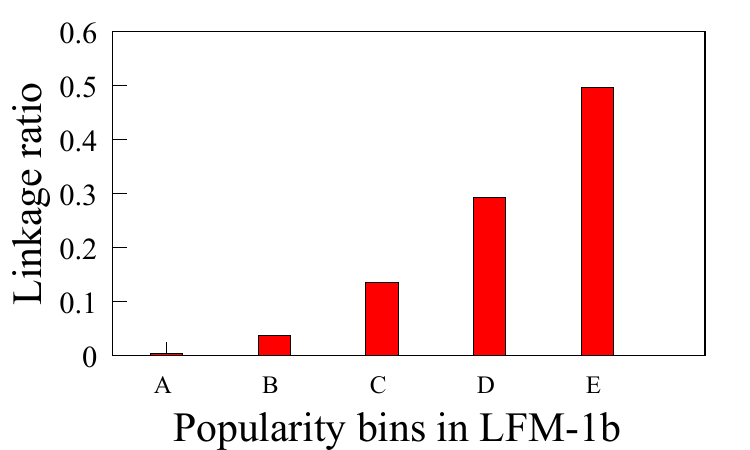}
\end{minipage}%
\begin{minipage}[t]{0.25\linewidth}
\centering
\includegraphics[width=1.60in]{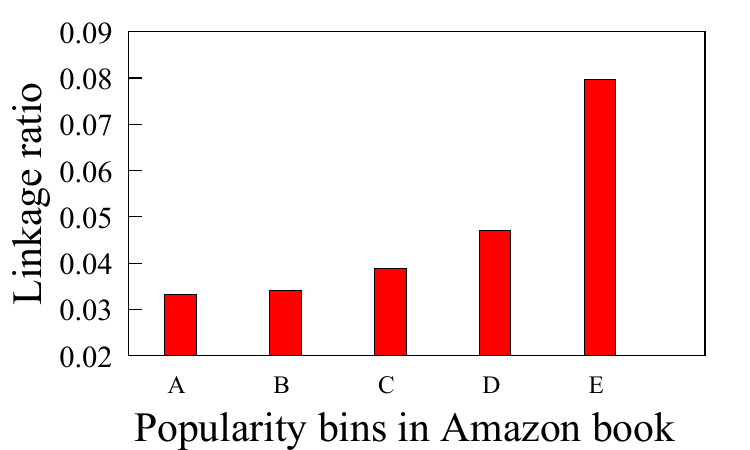}
\end{minipage}
}%
\subfigure[Recency.]{
\begin{minipage}[t]{0.25\linewidth}
\centering
\includegraphics[width=1.60in]{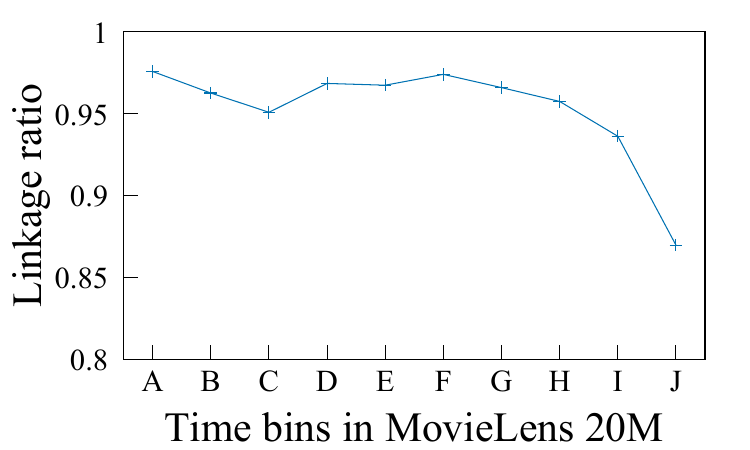}
\end{minipage}
}%
\centering
\caption{Examining the effect of two factors on the linkage results. We use $A$, $B$, $\cdots$ to indicate the bin number in an ordered way. The first three subfigures correspond to the popularity analysis, and the last one corresponds to the recency analysis. }\label{fig-popularity}
\end{figure*}

\paratitle{Effect of Popularity on Linkage}.
Intuitively, a popular RS item should be more likely to be included in a KB than an unpopular item, since
it is reasonable to incorporate more ``important" RS items judged by the RS users into KBs.
The construction of KB itself usually involves manual efforts, which is difficult to avoid the bias of human attention.
To measure the popularity of a RS item, we adopt a simple frequency-based method by counting the number of  users who have interacted with the item.
This measure characterizes the attractiveness of an item from the users in a RS. First, we sort the items ascendingly according to its popularity value.
Then, we further equally divide all the items into five ordered bins with the same number of items.
Hence, an item with a larger bin number will be more popular than another with a smaller bin number.
Then we compute the linkage ratio for each bin and the results are reported in Fig.~\ref{fig-popularity}(a) (the three subfigures on the left). It can be observed that  a bin with a larger number has a higher linkage ratio than the ones with a smaller number. The results indicate that popularity is likely to have positive effect on linkage.

\paratitle{Effect of Recency on Linkage}. The second factor we consider is the recency, \ie the time when a RS item was created.  Our assumption is that if a RS item was created or released on an earlier time, it would be more probable to be included in KBs.
Since human attention aggregation is a gradually growing process, a RS item usually requires a considerable amount of time to become popular.
To check this assumption, we need to obtain the release date of RS items. However, only the MovieLens $20M$ dataset contains such an attribute information, we only report the analysis result on this dataset. We first sort the items according to their release dates ascendingly, and then equally divide all the items into ten ordered bins following the procedure of the above popularity analysis.
Finally, we compute the linkage ratios for each bin. The results are reported in Fig.~\ref{fig-popularity}(b).
We can see that the linkage ratios gradually decrease with time going. The results indicate that recency is likely to have negative effect on linkage, \ie an older RS item seems to be more probable to be included in a KB than a more recent one. Especially, the last bin has a dramatic drop.
A possible reason is that our version of MovieLens is April 2015.


\section{Experiment}
In this section, we present the comparison of some existing recommendation algorithms on our linked datasets.

\paratitle{Experimental Setup.} Since our linked datasets are very large,
we first generate a small test set for the following experiments. 
We take the subset from the last year for LFM-1b dataset and the subset from year 2005 to 2015 for MovieLens $20M$ dataset.
We also perform 3-core filtering for Amazon book dataset and 10-core filtering for other datasets. The statistics of dataset used in \cite{HuangZDWC18} is reported in Table~\ref{tab-link-data} (the last column).
Following~\cite{He-IJCAI-2018}, we consider the last-item recommendation task for evaluation.
We set up such a task since it is a commonly used evaluation setting for RSs, and it is easy  to compare different methods. 
Given a user, first we sort the items according to the interaction timestamp ascendingly, then
we take the last item into the test set and the rest into training set. The final goal is to predict
the last item given the previous interaction sequence of a user. Since enumerating all the
items as candidate is time-consuming, we pair each ground-truth with 100 negative items to form a
randomly ordered list. Then each comparison method is to return a ranked list according its recommendation confidence. To evaluate different methods, we adopt a variety of evaluation metrics, including the Mean Reciprocal
Rank (MRR), Hit Ratio (HR), and Normalized Discounted cumulative gain (NDCG).

\paratitle{KB Information Representation.} Our focus is to provide rich knowledge information for recommender systems.
A simple way is to represent KB information with a one-hot vector, which is sparse and
large. Here we borrow the idea in \cite{KDD-2016-Zhang,TransE} to  embed KB data into low-dimensional vectors. Then the learned embeddings are used for subsequent recommendation algorithms.  To
train TransE~\cite{TransE}, we start with linked entities as seeds and expand the
graph with one-step search. Not all the relations in KBs are useful,
we remove unfrequent relations with fewer than 5,000 triples. After that, each linked
item is associated with a learned KB embedding vector.

\paratitle{Methods to Compare.} We consider the following methods for performance comparison\footnote{Here, since our purpose is to illustrate the use of this linked dataset, we only select four methods for performance comparison. We will try more knowledge-ware recommendation algorithms as future work. }:
\begin{itemize}
\item \textbf{BPR}~\cite{Rendle-uai09}: It learns a matrix factorization model by minimizing the pairwise ranking loss in a Bayesian framework.
\item \textbf{SVDFeature}~\cite{Chen2012SVDFeature}: It is a model for feature-based collaborative filtering. In this paper we use the KB embeddings as context features to feed into SVDFeature.
\item \textbf{mCKE}~\cite{KDD-2016-Zhang}: It first proposes to incorporate KB and other information to improve the recommendation performance. For fairness, we implement a simplified version of CKE by only using KB information, and exclude image and text information. Different from the original CKE, we fix KB representations and adopt the learned embeddings by TransE.
\item \textbf{KSR}~\cite{HuangZDWC18}: It is a Knowledge-enhanced Sequential Recommender (KSR). It incorporates KB information to enhance the semantic representation memory networks.
\end{itemize}

\paratitle{Results and Analysis.}
The results of different methods for the last-item recommendation are presented in Table~\ref{tab:results-recommendation}. We can see that:

(1) Among all the methods, BPR performs worst on three datasets. Overall, the other models perform better than BPR, since they incorporate the information of KB. 

(2) SVDFeature is implemented with a pairwise ranking loss function, and it can be roughly understood as an enhanced BPR model with the incorporation of the learned KB embeddings. Compared with BPR, SVDFeature is slightly better on
the book dataset which is more sparse, and substantially better on the music dataset and the book dataset. In SVDFeature,
each additional context feature will increase some number of parameters (deciding on the number of KB embeddings dimensions). 

(3) Next, we analyze the performance of the knowledge-aware recommendation methods, namely mCKE and KSR.
Overall, mCKE does not work well as expected, which only beats SVDFeature on the LFM-1b dataset.
A possible reason is that our implementation of mCKE fixes the learned KB embeddings, while the original CKE model adaptively updates KB embeddings.
As a comparison, the recently proposed KSR method works best consistently on the three datasets.
KSR combines the capacity of modeling data sequences from Recurrent Neural Networks (RNN) and the capacity of storing data in a long term from Memory Networks (MN). It further enhances MNs with the learned KB embeddings.

\begin{table}
	\centering
	\begin{scriptsize}
		\caption{Performance comparison of different methods on the task of last-item recommendation.}
		\label{tab:results-recommendation}%
		\begin{tabular}{|l r|l|l c c |}
			\hline
			Datasets& &Methods&MRR&Hit@10&NDCG@10
            \\
           		\hline \hline
			MovieLens $20M$ & & BPR &0.128&0.276&0.144\\
			~ ~\#\emph{users}: &61,583& SVDFeature &0.204&0.448&0.243\\
			~ ~\#\emph{items}: &19,533& mCKE &0.178&0.382&0.209 \\
			~ ~\#\emph{interactions}: &5,868,015& KSR &0.294&0.571&0.344\\
			\hline
            LFM-1b & &BPR &0.227&0.458&0.265\\
			~ ~\#\emph{users}:&7,694& SVDFeature &0.337&0.544&0.373\\
			~ ~\#\emph{items}:&30,658& mCKE &0.371&0.541&0.399 \\
			~ ~\#\emph{interactions}&203,975& KSR &0.427&0.607&0.460\\
			\hline
           Amazon book& &BPR &0.222&0.505&0.272\\
			~ ~\#\emph{users}: &65,125& SVDFeature &0.264&0.544&0.315\\
			~ ~\#\emph{items}: &69,975& mCKE &0.248&0.494&0.291 \\
			~ ~\#\emph{interactions}: &828,560& KSR &0.353&0.653&0.413\\
			\hline
		\end{tabular}%
	\end{scriptsize}
\end{table}%

\section{Conclusion}
This paper introduced a public dataset for linking RS with KB, namely \emph{KB4Rec  v1.0}. Our dataset covered three domains consists of a large number of linked ID pairs.
As future work, we will consider linking more RS datasets with Freebase.
We will also consider adopting other KB data for linkage, \eg YAGO and DBpedia.


\bibliographystyle{ACM-Reference-Format}
\bibliography{main}

\end{document}